\documentclass[prl, amsmath,twocolumn, amssymb,superscriptaddress, floatfix]{revtex4-1}

\usepackage{amsmath}
\usepackage{amssymb}
\usepackage{amsthm}
\usepackage[pdftex]{color}
\usepackage{graphicx}
\usepackage{dcolumn} 
\usepackage{bm} 
\usepackage{epic}
\usepackage{longtable}
\usepackage{ulem}   
\newcommand{\HH}{\mathcal{H}}
\newcommand{\kk}{\textbf{k}}
\newcommand{\p}{\textbf{p}}

\newcommand{\mT}{\mathcal{T}}
\newcommand{\MT}{\mathcal{T}_z}
\newcommand{\mI}{\mathcal{I}}

\begin{document}
\title{
    Superconductivity without inversion and time-reversal symmetries
      }
\author{Mark H.~Fischer}
\affiliation{Institute for Theoretical Physics, ETH Zurich, 8093 Zurich, Switzerland}
\author{Manfred Sigrist}
\affiliation{Institute for Theoretical Physics, ETH Zurich, 8093 Zurich, Switzerland}
\author{Daniel F.~Agterberg}
\affiliation{Department of Physics, University of Wisconsin-Milwaukee, Milwaukee, WI 53201}
\date{\today}

\begin{abstract}

 The traditional symmetries that protect superconductivity are time-reversal and inversion.
 Here, we examine the minimal symmetries protecting superconductivity in two dimensions and find that time-reversal symmetry and inversion symmetry are not required, and having a combination of either symmetry with a mirror operation on the basal plane is sufficient. We classify superconducting states stabilized by these two symmetries, when time-reversal and inversion symmetries are not present, and provide realistic minimal models as examples. Interestingly, several experimentally realized systems, such as transition metal dichalcogenides and the two-dimensional Rashba system belong to this category, when subject to an applied magnetic field.
\end{abstract}

\maketitle

\textit{Introduction -- }%
Progress in the fabrication of atomically thin material layers including ones showing superconductivity~\cite{novoselov:2016, uchihashi:2017} invites a new look at the question of stability and classification of superconducting states in two dimensions (2D).
In a three-dimensional system, inversion $\mI$ and time-reversal symmetry (TRS) $\mT$ are the only symmetries that ensure that a fermionic state with momentum $\vec{k}$ is (energy) degenerate with another state at $-\vec{k}$ \cite{anderson:1984}.  This degeneracy is necessary to support a weak-coupling superconducting instability due to the formation of Cooper pairs with total momentum $\vec{q}=0$. Consequently, these two symmetries are at the basis of the usual classification of even/spin-singlet and odd/spin-triplet Cooper pairs~\cite{sig91}. However, in 2D, the situation is different because momentum $\kk = (k_x, k_y)$ lies only in-plane. In addition to $\mI$ and $\mT$,  $M_z \mI=C_{2z}$ and $M_z \mT\equiv \MT$, with $M_z$ the mirror operation relating $z$ to $-z$, also ensure degeneracy between states with momenta $\kk$ and $-\kk$. 
Being outside the standard classification, these two additional symmetries could give rise to novel superconducting states.

In fact,  the two symmetries are essential for recent developments in 2D superconductors lacking $\mT$ and $\mI$: the well-known platform for creating Majorana edges states in 2D, a metallic electron gas with Rashba spin-orbit coupling and an out-of-plane magnetization, remains superconducting because this configuration preserves $C_{2z}$ symmetry. In addition, monolayer transition metal dichalcogenides (TMDs) retain $\MT$ in an applied in-plane Zeeman field allowing Cooper pairs to be formed, unhindered by the magnetic field. This  robustness of the superconducting state to in-plane fields has indeed been observed~\cite{lu:2015, xi:2015, saito:2015,delabarrera:2017tmp}.
Another example is provided by the recent proposal of (N\'eel) antiferromagnetic order in superconducting monolayer FeSe~\cite{coh:2015, wang:2016}.  This provides an intriguing example of a 2D system without $\mT$ and $\mI$, but where both $C_{2z}$ and $\MT$ are conserved. 
Importantly, the absence of both $\mI$ and $\mT$ in all these systems suggests that usual Cooper pairs are suppressed, leading naturally to the question if the remaining $C_{2z}$ and/or $\MT$ symmetries give rise to new superconducting states.

In this work, we answer this question in the affirmative. 
We initially present a topological classification of 2D superconductors without $\mT$ or $\mI$ by formulating the usual theory of superconductivity in terms of $\MT$ and $C_{2z}$ ~\cite{bzdusek:2017, sato:2017}. We then provide microscopic models that belong to each of these classes allowing us to discuss the most probable order parameters in each case as characterized by the notion of superconducting fitness~\cite{fischer:2013b, ramires:2016}. Specifically, we focus on the three aforementioned examples: superconductivity (a) with Rashba spin-orbit coupling and a $z$-axis magnetization (only $C_{2z}$), (b) in TMDs with an in-plane magnetic field (only $\MT$), and (c) in the N\'eel-ordered monolayer FeSe (both $C_{2z}$ and $\MT$).  
 
\begin{table}[b]
\centering
\begin{tabular}{c|c|c|c|c|c|c|c}
  \hline
  \hline
  nodal class & $\MT$ & $C_{2z}$ & $\mathfrak{T}$ & $\mathfrak{P}$ & $\mathfrak{C}$ & nodes & bulk class \\
  \hline
  C & $\times$ &  even & $\times$ & $-1$ & $\times$ &  nodeless& D ($\mathbb{Z}$)\\
  AIII & $1$ & $\times$ & $\times$ & $\times$ & $1$ &  nodal  ($\mathbb{Z}$)& BDI\\
  CII & $1$ & even & $-1$ & $-1$ & $1$ & nodeless & BDI\\
  DIII & $1$ & odd & $-1$ & $1$ & $1$ &  nodal ($2\mathbb{Z}$) & BDI\\
  \hline
  \hline
\end{tabular}
\caption{Summary of our main findings for superconductivity in a 2D system lacking inversion and TRS, but possessing $\MT = M_z \mT$ and/or $C_{2z}$. The symmetries that relate $\kk$ to $\kk$ are $\mathfrak{T} = C_{2z} \mathcal{T}_z$, $\mathfrak{P} = C_{2z} P$, $\mathfrak{C} = \mathfrak{TP}$, and $P$ is the usual particle-hole symmetry. In the last two columns, the topological charge of the nodes (bulk) is also given for the case of protected gap nodes (non-trivial bulk / edge states). With $\MT^2=1$, the bulk can only be non-trivial for broken $\mathcal{T}_z$ in 2D.}
\label{tab:sum}
\end{table}

\textit{Topological classification -- }%
Prior to examining the specific examples, it is worthwhile making some general statements about pairing when just the symmetries $C_{2z}$ or $\MT$ are present. We consider two relevant issues: 1) Can such superconductors be nodal? 2) If they are fully gapped, can the bulk be topologically non-trivial and Majorana edge states appear? Table~\ref{tab:sum} summarizes the results of our analysis. To address the possibility of nodes, we follow the nodal AZ+$\mI$ classification of Ref.~\onlinecite{bzdusek:2017}. In particular,  we use symmetries local in ${\bf k}$ space to classify nodes. Here, the relevant anti-unitary symmetries are $\mathfrak{T}=C_{2z}\MT $, $\mathfrak{P}=C_{2z}P$ with $P$ the particle-hole symmetry, as well as the unitary chiral symmetry stemming from the product $\mathfrak{C} = \mathfrak{T}\mathfrak{P}$. They operate on the BdG Hamiltonian as follows
\begin{eqnarray}
\mathfrak{T}H(\kk)\mathfrak{T}^{-1}=&&H(\kk) \nonumber\\
\mathfrak{P}H(\kk)\mathfrak{P}^{-1}=&&-H(\kk)\nonumber\\
\mathfrak{C} H(\kk) \mathfrak{C}^{-1}=&& -H(\kk).
\end{eqnarray}
These symmetries allow for the topological classification of nodal charges, with the nodal classes defined in the traditional way~\cite{ryu:2010}. Note the symmetries are labeled by $1,-1$ or $\times$, where the label $\times$ indicates the absence of the symmetry and $\pm 1$ correspond to $\mathfrak{T}^2=\pm 1$ and 
$\mathfrak{P}^2=\pm 1$. Finally, for the chiral symmetry $\mathfrak{C} $, the label $1$ indicates this symmetry exists. We will elaborate on the results of Tab.~\ref{tab:sum} through the discussion of the microscopic models.  

For the fully gapped situation, we can identify the conditions required for topologically non-trivial superconductors with Majorana edge states~\cite{ryu:2010}. We do not consider the role of $C_{2z}$, as this symmetry is broken by all edges and is therefore not relevant for Majorana edge states. The resultant classification depends only upon the presence or absence of $\MT$ symmetry. When this symmetry exists, $\MT^2=1$, and the relevant fully gapped class is BDI, which does not allow for Majorana edges states in 2D  \cite{ryu:2010}. Without $\MT$ present, the class is D, which allows for Majorana edges states protected by a $\mathbb{Z}$ invariant  \cite{ryu:2010}.

\textit{Only $C_{2z}$ present -- }%
In this case the nodal topological class is C (see Tab.~\ref{tab:sum}). It is instructive to understand how this arises.
With just $C_{2z}$ present, the bands are non-degenerate and we focus in the following on a single band with dispersion $\xi_{\kk}$. A Cooper pair formed in such a non-degenerate band can be written as $\langle \kk ; (C_{2z}\!:\!\kk)\rangle$, where $(C_{2z}\!:\!\kk)$ refers to the state obtained when applying $C_{2z}$. Using $C_{2z}^2 = -1$ and Fermi statistics, it immediately follows that any resulting order parameter $\Delta_\kk$ is even under $C_{2z}$. It is this constraint that implies that such superconducting states belong to the nodal class C and are generally nodeless \cite{schnyder:2015, chiu:2016}.

In the BdG basis corresponding to the Hamiltonian
\begin{equation}
    H(\kk) = \begin{pmatrix} \xi_{\kk} & \Delta_\kk\\ \Delta^*_{\kk} & - \xi_{-\kk}\end{pmatrix}
\end{equation}
the $C_{2z}$ operator reads
\begin{equation}
C_{2z}=\left(\begin{array}{cc}
                     e^{i\phi_\kk} & 0 \\
                     0 & e^{-i\phi_{-\kk}} \\
                   \end{array}
                 \right)
\end{equation}
with $e^{i(\phi_\kk+\phi_{-\kk})}=-1$ and $H(\kk)$ satisfies $C_{2z}H(\kk) C_{2z}^{\dagger}=H(-\kk)$.
Particle-hole symmetry is implemented as $P=\tau_x K$, where $K$ is the complex conjugation operator and $\tau_i$ denotes a Pauli matrix.
This satisfies the relation  $PH(\kk)P^{-1}=-H(-\kk)$. Finally, $\mathfrak{P}$ operates as $\mathfrak{P}H(\kk)\mathfrak{P}^{-1}=-H(\kk)$ and satisfies $\mathfrak{P}^2=-1$, while the system possesses neither $\mathfrak{T}$ nor $\mathfrak{C}$.  

With $\mathfrak{P}^2=-1$, we can choose a basis in which $\mathfrak{P}=i\tau_yK$. Parametrizing $H(\kk)=\sum_i c_i(\kk) \tau_i$, $\mathfrak{P}$  implies only that $c_0(\kk)=0$. Hence, we find a dispersion $\pm\sqrt{c_x(\kk)^2+c_y(\kk)^2+c_z(\kk)^2}$, for which to vanish we need three constraints. As this is not generally possible in two-dimensions, we conclude this dispersion must be fully gapped. 

For this case, fully gapped superconductors are defined by just $P^2=1$, putting them in class D yielding a familiar integer Chern number in 2D. This also gives rise to Majorana modes, which can be non-Abelian, as is well known from a 2D Rashba $s$-wave superconductor in a $z$-axis Zeeman field~\cite{sato:2009}. 

We proceed with a specific example to examine the microscopic gap structure. We consider the simplest model of a two-dimensional system violating both $\mI$ and $\mT$ described by the (normal-state) Hamiltonian
\begin{equation}
    \hat{\HH}_{\kk} = \varepsilon_{\kk}\hat{\sigma}^0 + (\vec{f}_{\kk} + \vec{m})\cdot\vec{\sigma},
    \label{eq:rashba}
\end{equation}
with $\hat{\sigma}^0$ and $\vec{\sigma}$ the identity and Pauli matrices, respectively, acting in spin space. Note that $\vec{f}_{\kk} = - \vec{f}_{-\kk} \neq 0$ parameterizes a spin-orbit coupling and can only be present when $\mI$ is lacking, while any magnetization $\vec{m}\neq0$ implies broken TRS. To enforce $C_{2z}$ symmetry and to have $\MT$ broken we choose  both $f^x_{\kk}$ and $f^y_{\kk}$ non-zero, with $f^x_{\kk}=\alpha k_y$ and $f^y_{\kk} = -\alpha k_x$ in the standard Rashba case, and the magnetization to point along the $z$ direction. Later, we consider $\MT$ preserved and $C_{2z}$ broken, for which  $f^z_{\kk}\neq 0$ and the magnetization is in-plane.

To analyze which pairing states are the most stable in the weak-coupling case we first employ the recently developed concept of order-parameter `fitness' as given by intra-band pairing~\cite{fischer:2013b, ramires:2016}.
According to this criterion, the most stable (weak-coupling) superconducting order parameters in spin space have to satisfy
\begin{equation} 
    [\hat{\HH}_{\kk},\hat{\Delta}_{\kk}]^*=\hat{\HH}_{\kk}\hat{\Delta}_{\kk}-\hat{\Delta}_{\kk}\hat{\HH}^{*}_{-\kk}=0.
    \label{eq:fitness}
\end{equation}
It is convenient to start from the stable order parameters without magnetic order, i.e., $m_z = 0$, which are a spin-singlet combined with a spin-triplet state with $\vec{d}_{\kk}\parallel \vec{f}_{\kk}$. Analyzing the stability of these order parameters against a magnetic field, we find that the spin-singlet gap is suppressed, while the spin-triplet gap remains unaffected.
This result, namely the stability of a triplet order parameter with a $d$ vector both parallel to $\vec{f}_{\kk}$ and perpendicular to the magnetization, is more general~\cite{sigrist:2007, smidman:2017} and also holds for the case of only $\MT$.

The condition Eq.~\eqref{eq:fitness} is equivalent to  considering a general gap function in the band basis and imposing that, for a weak coupling instability, the pairing is purely intra-band. Having gap magnitudes $\Delta_1$ and $\Delta_2$ on the two non-degenerate bands of the Hamiltonian Eq.~\eqref{eq:rashba}, the gap in the original spin basis reads~\cite{suppmat}
\begin{multline}
    \hat{\Delta} = \{\Delta_-(\hat{f}_\kk^x \hat{\sigma}^x + \hat{f}_\kk^y \hat{\sigma}^y) + \\ 
    \Delta_+ [\sin\vartheta_\kk\hat{\sigma}^0 - i \cos\vartheta_\kk(\hat{f}_\kk^x \hat{\sigma}^y - \hat{f}_\kk^y \hat{\sigma}^x)]\}(i\hat{\sigma}^y),
    \label{eq:c2v_full_delta}
\end{multline}
with $\Delta_{\pm} = (\Delta_1\pm\Delta_2)/2$, $\tan \vartheta_{\kk} = |\vec{f}_{\kk}| / |\vec{m}|$, and $\hat{f}^i_{\kk} = f^i_{\kk} / |\vec{f}_{\kk}|$.
A general order parameter thus consists of the above discussed stable triplet component $\Delta_{-}$, a singlet component, and finally a triplet component with a $d$ vector perpendicular to both the magnetic field and the spin-orbit vector. 
The triplet component $\Delta_{-}$ with $\vec{d}_{\kk} \parallel \vec{f}_{\kk}$ corresponds to a helical state, i.e., two chiral states of equal magnitude but opposite chirality for up and down spins, and thus results in a spin current but no net charge current at a boundary. The additional triplet component, however, breaks the balance between the two spin orientations and hence, results in a finite charge current~\cite{suppmat}.

Two remarks are in order before continuing regarding the combination of the latter two terms with magnitude $\Delta_{+}$ in Eq.~\eqref{eq:c2v_full_delta}. First, the additional spin-triplet part is a pure inter-band order parameter for large spin-orbit coupling in the absence of magnetization. Second, the exact combination of these two terms also describes a pure intra-band order parameter and indeed satisfies Eq.~\eqref{eq:fitness}.
However, this implicitly assumes a pairing interaction leading to this fine-tuned combination, which is not true in general. Consequently, the resulting self-consistent pairing state will not satisfy Eq.~\eqref{eq:fitness}, but still have the three components as in Eq.~\eqref{eq:c2v_full_delta}. In the following, we thus restrict ourselves to the fitness analysis of stability against a magnetic field or magnetic order.

\begin{figure}[t]
    \centering
    \includegraphics[width=0.45\textwidth]{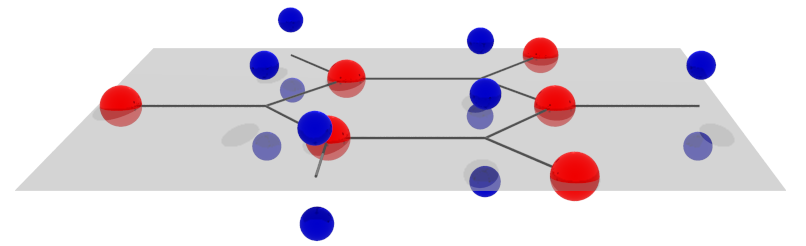}
    \caption{Schematic crystal structure of a single layer TMD with the formula MX$_2$ lacking inversion symmetry. Adding an in-plane field additionally breaks time-reversal symmetry, while conserving $\MT$.}
    \label{fig:tmd}
\end{figure}

\textit{Only $\MT$ present --} Here, the nodal class is CII with stable point nodes. This situation is realized when applying an in-plane magnetic field to monolayer TMDs, which have the chemical formula MX$_2$~\cite{xiao:2012}, with M a transition metal and X a chalcogen, see Fig.~\ref{fig:tmd}. While the three-fold rotation-symmetric crystal structure breaks inversion symmetry, when adding an in-plane magnetic field the system additionally lacks time-reversal symmetry, but retains $\MT$. This could be achieved by sandwiching the layer between ferromagnets with in-plane magnetization or when applying an in-plane field.
Indeed, exceptionally high critical in-plane fields $H_{c2}$ were observed in superconducting TMDs and ascribed to their Ising-type superconductivity~\cite{lu:2015, xi:2015, saito:2015,delabarrera:2017tmp}. Here, this Ising superconductivity is readily understood as stemming from a large $f_{\kk}^z$ locking the spins along the $z$ direction. Due to the remaining $\MT$ symmetry, an in-plane field will not destroy superconductivity. As mentioned above, there is a fully stable triplet order parameter with a $d$ vector only having a $z$ component~\cite{suppmat}. 

With only $\MT$ symmetry present, the system has neither $\mathfrak{T}$ nor $\mathfrak{P}$ symmetries. However, it has a chiral symmetry $\mathfrak{C} = \mathfrak{T}\mathfrak{P}=\MT P$, which provides the same protection of point nodes as the known $C=TP$~\cite{schnyder:2011, schnyder:2012}. This new chiral symmetry extends the topological protection to situations in which an in-plane magnetic field is applied, provided $\MT$ is conserved. Hence, this allows for the existence of protected point nodes with a $\mathbb{Z}$ classification and corresponding Majorana flat-band edge states. 

We can further classify the fully gapped superconducting states using the AZ classification. Replacing $\mT$ with $\MT$ yields $(\MT)^2=1$, such that the states do not fall into the usual DIII superconducting class (with a Kane-Mele $\mathbb{Z}_2$ charge in 2D), but rather into class BDI, without topological charges and hence, no protected edge states in 2D.

\textit{$\MT$ and $C_{2z}$ present -- }%
In this case, the nodal topological class is either CII or DIII depending on whether the gap function is even or odd under $C_{2z}$ symmetry.
The three order parameters even under $C_{2z}$ 
correspond to Cooper pairs of the form $\langle \kk; (\MT\!:\!\kk)\rangle - \langle (C_{2z} \MT\!:\!\kk); (C_{2z}\!:\!\kk)\rangle$, $\langle \kk; (C_{2z}\!:\!\kk)\rangle$, and $\langle (C_{2z}\MT\!:\!\kk); (\MT\!:\!\kk)\rangle$, while the odd one corresponds to $\langle \kk; (\MT\!:\!\kk)\rangle + \langle (C_{2z} \MT\!:\!\kk); (C_{2z}\!:\!\kk)\rangle$ with nodes classified by the same chiral symmetry discussed above. In contrast to the case of inversion and time-reversal symmetry, these Cooper pairs are in general not pseudospin triplet and pseudospin singlet.
Note that presence of $\MT$ symmetry again implies that no Majorana-modes appear for fully gapped superconducting states.

\begin{figure}[b]
    \centering
    \includegraphics[width=0.45\textwidth]{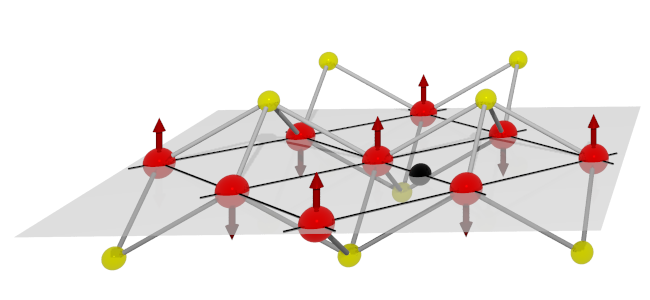}
    \caption{Crystal structure and N\'eel order of the FeSe single layer with Fe (red) and Se (yellow). The black lines indicate the Fe square lattice and the black dot denotes the inversion center without N\'eel order in between Fe sites.}
    \label{fig:fese}
\end{figure}

Next, we turn to a microscopic model that provides representative examples of these novel states.
The previously introduced  single-band microscopic model necessarily breaks either $\MT$ or $C_{2z}$. In order to discuss a combination of the two, we thus turn to the example of N\'eel-ordered monolayer FeSe as depicted in Fig.~\ref{fig:fese}.
Monolayer FeSe exhibits superconductivity with greatly increased critical temperature compared to their bulk counterparts with reported $T_c$ up to $109$K~\cite{huang:2017}. While bulk FeSe shows no magnetic order~\cite{mcqueen:2009}, there is ongoing debate on such order in the monolayer, with proposals including N\'eel order~\cite{coh:2015, wang:2016}, as well as collinear antiferromagnetic order as in other iron pnictides~\cite{liu:2012c, cao:2014} and pair-checkerboard antiferromagnetic order~\cite{cao:2015}.

The monolayer FeSe has the same (non-symmorphic) space group symmetry as bulk iron pnictides, with a point group isomorphic to $D_{4h}$~\footnote{Note that in principle, the existance of a substrate will break $M_z$. However, as the substrate is not (strongly) chemically coupled to FeSe~\cite{liu:2012c}, we neglect this additional symmetry breaking here.}.
Since there are two Fe sites per unit cell, the standard N\'eel order has the same translation symmetry as the underlying lattice.
However, the system lacks time-reversal symmetry and inversion symmetry.
Importantly, as we consider the moments oriented along the $z$ direction (see Fig.~\ref{fig:fese}), the symmetry group still contains both $C_{2z}$ and $\MT$~
\footnote{Note that $\MT$ contains a non-trivial lattice translation inherited from $M_z$.}, the elements stabilizing a weak-coupling instability. 

Unlike most iron-based superconductors, the normal state has only electron-like bands around the $M$ points in the Brillouin zone crossing the Fermi energy~\cite{liu:2012d}. 
We can thus construct a $\kk\cdot\p$-type Hamiltonian working in a two-orbital basis and using Pauli matrices $\hat{\tau}^i$ operating in orbital space and $\hat{\sigma}^i$ operating in spin space.
The symmetry restricted Hamiltonian, i.e., invariant under the operations of the point group~\cite{cvetkovic:2013}, without magnetic order thus reads~\cite{agterberg:2017}
\begin{equation}
    \hat{\HH}_{\kk}=(\epsilon_0+a_1k_xk_y\hat{\tau}^z)\hat{\sigma}^0+a_2 \hat{\tau}^x(k_x\hat{\sigma}^y+k_y\hat{\sigma}^x). 
  \label{eq:Hfese}
\end{equation}
Microscopically, the first two terms correspond to normal hopping processes, while the last term stems from a staggered Rashba-type spin-orbit coupling~\cite{fischer:2011b}. The excitation spectrum 
given by Eq.~\eqref{eq:Hfese} can be well fitted to ARPES data~\cite{agterberg:2017}. In this basis, the main symmetry operations relating $\kk$ to $-\kk$ are given by $C_{2z}=i\hat{\tau}^0\hat{\sigma}^z$, $\mT=i\hat{\tau}^0\hat{\sigma}^y K$, and $\mI=\hat{\tau}^z\hat{\sigma}^0$. 

The reduction of the symmetry due to the magnetic order  allows for additional terms in the Hamiltonian,
\begin{equation}
    \delta \hat{\HH}_{\kk}=M_m\hat{\tau}^x \hat{\sigma}^z+\epsilon M_m (k_x^2-k_y^2) \hat{\tau}^y\hat{\sigma}^0.
    \label{eq:dHfese}
\end{equation}
Note that there are no additional operators of the form $\hat{\tau}^i\hat{\sigma}^j$, even if higher powers of $\kk$ are included. The first term in the above equation corresponds to a direct exchange term (Zeeman), while from a microscopic point of view, the second term stems from a 4th-nearest-neighbor hopping~\cite{suppmat} and can thus be expected to be minuscule. 
These new terms together with $\hat{\HH}_{\kk}$ lead to a gapped Dirac spectrum, which, however, is still doubly degenerate due to the combination of $C_{2z}$ and $\MT$. 

Following our above discussion of the single-band case, we analyze the weak-coupling pairing instabilities starting from the order parameters that are most stable without magnetic order. In particular,  
\begin{eqnarray}
    \hat{\Delta}_1({\bf k})=&& \psi_1({\bf k})\hat{\tau}^0 i \hat{\sigma}^y\label{eq:intrasinglet},\\
    \hat{\Delta}_2({\bf k})=&& \psi_2({\bf k})\hat{\tau}^0(k_x\hat{\sigma}^y+k_y\hat{\sigma}^x)i \hat{\sigma}^y\label{eq:intratriplet},\\
    \hat{\Delta}_3({\bf k})=&&\psi_3({\bf k})\hat{\tau}^z(k_x\hat{\sigma}^x-k_y\hat{\sigma}^y)i\hat{\sigma}^y \label{eq:interinplanetriplet},\\
    \hat{\Delta}_4({\bf k})=&&d_z({\bf k})\hat{\tau}^z\hat{\sigma}^z i\hat{\sigma}^y\label{eq:interdz},
\end{eqnarray}
where $\psi_i({\bf k})$ is an even function of ${\bf k}$ and $d_z({\bf k})$ is an odd function of ${\bf k}$.
These order parameters describe an intra-sublattice spin-singlet, Eq.~\eqref{eq:intrasinglet}, an intra-sublattice spin-triplet, Eq.~\eqref{eq:intratriplet}, as well as two inter-sublattice spin-triplet gap functions, Eqs.~\eqref{eq:interinplanetriplet} and \eqref{eq:interdz}~\cite{suppmat}. 
The former three, i.e.,  $\hat{\Delta}_1(\kk)$, $\hat{\Delta}_2(\kk)$, and $\hat{\Delta}_3(\kk)$, are $C_{2z}$ even and thus belong to nodal class CII. $\hat{\Delta}_4({\bf k})$, on the other hand, is odd and thus, presents an example of nodal class DIII, which can have topologically protected gap nodes.

Once the N\'eel order is added, only $\hat{\Delta}_4({\bf k})$ remains completely `fit'. 
There are, however, further order parameters that might be stable, if certain terms in the Hamiltonian are small. 
Notably, the spin-singlet gap function $\hat{\Delta}({\bf k})=\Delta_0 \hat{\tau}^z i\hat{\sigma}^y$ has been argued to be relevant for ``small'' spin-orbit coupling~\cite{agterberg:2017,shishidou:2018} -- namely the $a_2$ term in Eq.~\eqref{eq:Hfese} or  the $\epsilon$ term  in Eq.~\eqref{eq:dHfese}~\cite{suppmat}. This order parameter, with $d_{xy}$ pairing symmetry due to its orbital structure, is unaffected by the presence of N\'eel order and might thus survive. 
The usual spin-singlet state, however, is strongly suppressed by the magnetic order, illustrating that (phonon-mediated) $s$-wave superconductivity is unlikely to survive in a magnetically ordered phase. 

For real $d_z(\kk)$, the state $\hat{\Delta}_4(\kk)$ has protected point nodes, as mentioned above. Such a (real) order parameter could be selected by adding in-plane uniaxial strain or nematic order.
Otherwise, to avoid nodes $\hat{\Delta}_4(\kk)$ will likely break (global) TRS. The resulting state with $d_z(\kk) \sim k_x + i k_y$ only retains $C_{2z}$ and thus, the state with spontaneously broken $\mT$ provides another example of fully gapped  class D with Majoranas edge states. 

Finally, the order parameters given in Eqs.~\eqref{eq:intrasinglet} and \eqref{eq:intratriplet} mix with other pairing components that have the same symmetry and similar to the case discussed for the Rashba system, this will result in `chiral' states. However, the states have opposite chirality on the two sublattices of the FeSe lattice~\cite{suppmat}. 
The state thus breaks TRS in the same way the magnetic order does. It is, however, topologically trivial, such that there is no protected edge states. For certain edges, e.g., the $[11]$ edge (diagonal in plane), this state should have edge currents.

\textit{Conclusions -- }%
We reexamined the question of superconductivity in 2D and the symmetries necessary to allow for a weak-coupling, i.e., pure intra-band-coupling, instability. In particular, we focused on systems, where inversion and TRS are broken -- the symmetries required in three dimensions -- but their combination with mirror symmetry on the basal plane is conserved. 

With just $\MT$ symmetry, relevant to TMDs with in-plane field, protected nodes can exist leading to Majorana flat-band edge states. 
We further identified the combination of particle-hole symmetry and $\MT$ as the chiral symmetry in 2D that protects the stability of nodes and associated edge states even in this case of (in-plane) magnetic fields. 
Interestingly, in strong fields the order parameter could become of dominant (nodal) spin-triplet character with edge states accessible to tunneling experiments.

With just $C_{2z}$ symmetry, relevant to the Rashba case with $z$-axis magnetization, the gap is generically fully gapped and allows for Majorana chiral edge states. 
Finally, with both symmetries, relevant to N\'eel order in FeSe, we find an energetically stable orbitally non-trivial chiral spin-triplet pairing state that also allows for Majorana chiral edge states. Our results thus reveal how the symmetries $C_{2z}$ and $\MT$ can play an crucial role in understanding 2D superconductors.

\begin{acknowledgements}
We thank Tom\'{a}\v{s} Bzdu\v{s}ek, Aline Ramires, Tatsuya Shishidou, and Mike Weinert  for valuable discussions. M.S. acknowledges support from the Swiss National Science Foundation and D.F.A. the hospitality of the Pauli Center of the ETH Zurich.
\end{acknowledgements}

\bibliography{refs}
\end{document}


\title{
    Supplementary Material: Superconductivity without inversion and time-reversal symmetries}
    
\begin{abstract}
In this supplemental material, we discuss a simple microscopic model for the case of only $C_{2z}$ or $M_z \mathcal{T}$ present
and provide a single-orbital tight-binding model for FeSe including an out-of-plane antiferromagnetic (AFM) order.
\end{abstract}

\maketitle
\section{$C_{2z}$: Rashba with $z$-axis magnetization}
We start our discussion by looking at a two-dimensional system without any magnetization, but lacking the mirror symmetry with respect to $z\rightarrow -z$, which has a general Hamiltonian 
\begin{equation}
    \hat{\HH} = \frac{\kk^2}{2m}\hat{\sigma}^0 + f_{\kk}^x \hat{\sigma}^x + f_{\kk}^y \hat{\sigma}^y,
    \label{eq:rashba}
\end{equation}
with $f_\kk^i$ an odd function of momentum. For the simplest Rashba case, these are given by $f_\kk^x= \alpha k_y$ and $f_\kk^y = -\alpha k_x$. 
Note that there is no $\hat{\sigma}^z$ term allowed if we require $C_{2z}$ and time-reversal symmetry.

We can straight-forwardly solve for the eigenfunctions and eigenenergies of the Rashba Hamiltonian~\eqref{eq:rashba} by performing a rotation in spin space around the axis perpendicular to the (in-plane) vector $(f_\kk^x, f_\kk^y)$, namely $(f_\kk^y, -f_\kk^x)$ by $\pi/2$, i.e., using the transformation
\begin{equation}
    \hat{U}_{\kk} = \frac{1}{\sqrt{2}} (\hat{\sigma}^0 - i \hat{f}_\kk^y \hat{\sigma}^x + i \hat{f}_\kk^x \hat{\sigma}^y) = \frac1{\sqrt{2}}\begin{pmatrix} 1 &  \hat{f}_\kk^x- i \hat{f}_\kk^y\\ - \hat{f}_\kk^x-i \hat{f}_\kk^y  & 1\end{pmatrix}.
    \label{eq:rashba_transformation}
\end{equation}
Here, we have introduced $\hat{f}_\kk^i = f_\kk^i / |f_\perp|$ and $|f_\perp| = \sqrt{(f_\kk^x)^2 + (f_\kk^y)^2}$ is the in-plane component of the spin part of the Hamiltonian.
This transformation leads to a diagonal Hamiltonian in terms of new states $|\alpha, \kk\rangle$, which are now not spin degenerate anymore. However, due to both time-reversal and $C_{2z}$, there is a degeneracy between states $|\alpha, \kk\rangle$ and $|\alpha, -\kk\rangle$, such that we can have a Cooper instability at $\q=0$ with an order parameter
\begin{equation}
    \Delta_\alpha(\kk) = \psi_\alpha(\kk) \sim \langle c_{\alpha, -\kk} c_{\alpha, \kk}\rangle,
    \label{eq:rashba_op}
\end{equation}
with operators $c^\dag_{\alpha, \kk}$ creating the states $|\alpha, \kk\rangle$.
Note that this pairing is independently possible for each band (from pure energetics of Fermi surface pairing). Finally, also note that $\psi_{\alpha}(\kk)$ is necessarily odd in momentum, since there is no additional degree of freedom to satisfy the fermionic statistics.

We can now use the transformation matrices given in Eq.~\eqref{eq:rashba_transformation} to find the gap function given in the original (spin) basis. 
It is easiest to do the transformation one band at a time, starting with the first one
\begin{eqnarray}
    \hat{\Delta}_1(\kk) &=&  U_{-\kk}^{\dag} \begin{pmatrix} \psi_1(\kk) & 0 \\ 0 & 0\end{pmatrix} U_{\kk}^* =  \frac{\psi_1(\kk)}{2}\begin{pmatrix} 1 &  \hat{f}_\kk^x - i \hat{f}_\kk^y\\ - \hat{f}_\kk^x-i \hat{f}_\kk^y  & 1\end{pmatrix} \begin{pmatrix} 1 & 0 \\ 0 & 0\end{pmatrix} \begin{pmatrix} 1 &  \hat{f}_\kk^x + i \hat{f}_\kk^y\\ - \hat{f}_\kk^x + i \hat{f}_\kk^y  & 1\end{pmatrix}\\
    &=& \frac{\psi_1(\kk)}2\begin{pmatrix} 1 & \hat{f}_\kk^x + i \hat{f}_\kk^y \\ -(\hat{f}_\kk^x + i \hat{f}_\kk^y) & - (\hat{f}_\kk^x + i \hat{f}_\kk^y)^2 \end{pmatrix}\\
    &=&  \frac{\Delta_1}2 \begin{pmatrix} \hat{f}_\kk^x - i \hat{f}_\kk^y & 1 \\ -1 & -\hat{f}_\kk^x - i \hat{f}_\kk^y \end{pmatrix} = \frac{\Delta_1}2 (\hat{\sigma}^0 - \hat{f}_\kk^x \hat{\sigma}^x - \hat{f}_\kk^y\hat{\sigma}^y)i\hat{\sigma}^y,
    \label{eq:delta1_ss}
\end{eqnarray}
where in the last line we used the simplest odd function $\psi_1(\kk) = \Delta_1(\hat{f}_\kk^x - i \hat{f}_\kk^y)$.
Similarly, for the second band we find
\begin{eqnarray}
    \hat{\Delta}_2(\kk) &=&  U_{-\kk}^{\dag} \begin{pmatrix} 0 & 0 \\ 0 & \psi_2(\kk)\end{pmatrix} U_{\kk}^* =  \frac{\psi_2(\kk)}{2}\begin{pmatrix} 1 &  \hat{f}_\kk^x - i \hat{f}_\kk^y\\ - \hat{f}_\kk^x-i \hat{f}_\kk^y  & 1\end{pmatrix} \begin{pmatrix} 0 & 0 \\ 0 & 1\end{pmatrix} \begin{pmatrix} 1 &  \hat{f}_\kk^x + i \hat{f}_\kk^y\\ - \hat{f}_\kk^x + i \hat{f}_\kk^y  & 1\end{pmatrix}\\
    &=& \frac{\psi_2(\kk)}2\begin{pmatrix} - (\hat{f}_\kk^x - i \hat{f}_\kk^y)^2 & \hat{f}_\kk^x - i \hat{f}_\kk^y \\ -(\hat{f}_\kk^x - i \hat{f}_\kk^y) & 1 \end{pmatrix}\\
    &=&  \frac{\Delta_2}2 \begin{pmatrix} - \hat{f}_\kk^x + i \hat{f}_\kk^y & 1 \\ -1 & \hat{f}_\kk^x + i \hat{f}_\kk^y \end{pmatrix} = \frac{\Delta_2}2 (\hat{\sigma}^0 + \hat{f}_\kk^x \hat{\sigma}^x + \hat{f}_\kk^y\hat{\sigma}^y)i\hat{\sigma}^y,
    \label{eq:delta2_ss}
\end{eqnarray}
where we have replaced $\psi_2(\kk) = \Delta_2(\hat{f}_\kk^x + i \hat{f}_\kk^y)$, such that we find the well-known result that the order parameter is a combination of singlet and triplet gap,
\begin{equation}
    \hat{\Delta}(\kk) = \frac12[(\Delta_1 + \Delta_2)\hat{\sigma}_0 + (\Delta_2 - \Delta_1)(\hat{f}_\kk^x \hat{\sigma}^x + \hat{f}_\kk^y \hat{\sigma}^y)](i\hat{\sigma}^y),
    \label{eq:rashba_full_delta}
\end{equation}
with the $d$ vector parallel to the spin-orbit vector $\vec{f}_{\kk}$.
For the Rashba case, the spin-triplet part of Eq.~\eqref{eq:rashba_full_delta} reads
\begin{equation}
    \hat{\Delta}^{(\rm t)}(\kk) \propto \begin{pmatrix} (\Delta_1 - \Delta_2)\frac{k_x - i k_y}{|\kk|} & 0 \\ 0 &  (\Delta_1 - \Delta_2)\frac{k_x + i k_x}{|\kk|} \end{pmatrix}
\end{equation}
i.e., it consists of a $p-ip$ ($p+ip$) order parameter for the spin up (down) electrons. This results in a spin current at the boundary, but no charge current.

We now add a magnetic field in the $z$ direction to the Hamiltonian given in Eq.~\eqref{eq:rashba},
\begin{equation}
    \hat{\HH} = \frac{\kk^2}{2m}\hat{\sigma}^0 + f_\kk^x \hat{\sigma}^x + f_\kk^y \hat{\sigma}^y + h_z \hat{\sigma}^z,
    \label{eq:c2z}
\end{equation}
and analyze how this changes the general order parameter given in Eq.\eqref{eq:rashba_full_delta}.
Note that the above Hamiltonian still preserves $C_{2z}$ symmetry, such that we can again find a superconducting instability with $\q=0$.

To diagonalize the Hamiltonian, we can use a similar spin rotation as before, however now we rotate by an angle $\vartheta_\kk$ with $\tan \vartheta_\kk = |f_\kk|/h_z$. The transformation matrices read accordingly
\begin{equation}
\hat{U}_{\kk} = [\cos\frac{\vartheta_\kk}2\hat{\sigma}^0 - i \sin\frac{\vartheta_\kk}2(\hat{f}_\kk^y \hat{\sigma}^x - \hat{f}_\kk^x \hat{\sigma}^y)] = \begin{pmatrix} \cos\frac{\vartheta_\kk}2 &  \tilde{f}_\kk^x- i \tilde{f}_\kk^y\\ - \tilde{f}_\kk^x-i \tilde{f}_\kk^y  & \cos\frac{\vartheta_\kk}2\end{pmatrix},
    \label{eq:c2z_transformation}
\end{equation}
where now $\tilde{f}_\kk^i = \sin(\vartheta_\kk/2)\hat{f}_\kk^i$.
We can follow the exact same protocol as above, namely for the first band
\begin{eqnarray}
    \hat{\Delta}_1(\kk) &=&  U_{-\kk}^{\dag} \begin{pmatrix} \psi_1(\kk) & 0 \\ 0 & 0\end{pmatrix} U_{\kk}^* =  \psi_1(\kk)\begin{pmatrix}\cos\frac{\vartheta_\kk}2  &  \tilde{f}_\kk^x - i \tilde{f}_\kk^y\\ - \tilde{f}_\kk^x-i \tilde{f}_\kk^y  & \cos\frac{\vartheta_\kk}2\end{pmatrix} \begin{pmatrix} 1 & 0 \\ 0 & 0\end{pmatrix} \begin{pmatrix} \cos\frac{\vartheta_\kk}2 &  \tilde{f}_\kk^x + i \tilde{f}_\kk^y\\ - \tilde{f}_\kk^x + i \tilde{f}_\kk^y  & \cos\frac{\vartheta_\kk}2\end{pmatrix}\\
    &=& \psi_1(\kk)\begin{pmatrix} \cos^2\frac{\vartheta_\kk}2& \cos\frac{\vartheta_\kk}2(\tilde{f}_\kk^x + i \tilde{f}_\kk^y) \\ -\cos\frac{\vartheta_\kk}2(\tilde{f}_\kk^x + i \tilde{f}_\kk^y) & - (\tilde{f}_\kk^x + i \tilde{f}_\kk^y)^2 \end{pmatrix}\\
    &=&  \Delta_1 \begin{pmatrix} \cos^2\frac{\vartheta_\kk}2(\hat{f}_\kk^x - i \hat{f}_\kk^y) & \frac12 \sin\vartheta_\kk \\ -\frac12 \sin\vartheta_\kk & -\sin^2\frac{\vartheta_\kk}2(\hat{f}_\kk^x + i \hat{f}_\kk^y) \end{pmatrix}\\
&=&  \frac{\Delta_1}2 [\sin\vartheta_\kk \hat{\sigma}^0 - (\hat{f}_\kk^x \hat{\sigma}^x + \hat{f}_\kk^y\hat{\sigma}^y) - i \cos\vartheta_\kk(\hat{f}_\kk^x \hat{\sigma}^y - \hat{f}_\kk^y\hat{\sigma}^x)]i\hat{\sigma}^y,
    \label{eq:c2z_delta1_ss}
\end{eqnarray}
with 
\begin{equation}
    \sin\vartheta_\kk = \frac{|\vec{f}_{\kk}|}{\sqrt{(f_\kk^x)^2 + (f_\kk^y)^2 + h_z^2}} \qquad {\rm and} \qquad \cos\vartheta_\kk = \frac{h_z}{\sqrt{(f_\kk^x)^2 + (f_\kk^y)^2 + h_z^2}}.
    \label{eq:angles}
\end{equation}
Note that we have used above that $\cos\vartheta_\kk = \cos \vartheta_{-\kk}$ and thus $\cos\frac{\vartheta_\kk}2 = \cos \frac{\vartheta_{-\kk}}2$ and $\sin\frac{\vartheta_\kk}{2} = \sin \frac{\vartheta_{-\kk}}2$. 
Again doing the same calculation for the second band, we find
\begin{eqnarray}
    \hat{\Delta}_2(\kk) &=&  U_{-\kk}^{\dag} \begin{pmatrix} 0 & 0 \\ 0 & \psi_2(\kk)\end{pmatrix} U_{\kk}^* =  \psi_2(\kk)\begin{pmatrix}\cos\frac{\vartheta_\kk}2  &  \tilde{f}_\kk^x - i \tilde{f}_\kk^y\\ - \tilde{f}_\kk^x-i \tilde{f}_\kk^y  & \cos\frac{\vartheta_\kk}2\end{pmatrix} \begin{pmatrix} 0 & 0 \\ 0 & 1\end{pmatrix} \begin{pmatrix} \cos\frac{\vartheta_\kk}2 &  \tilde{f}_\kk^x + i \tilde{f}_\kk^y\\ - \tilde{f}_\kk^x + i \tilde{f}_\kk^y  & \cos\frac{\vartheta_\kk}2\end{pmatrix}\\
    &=& \psi_2(\kk)\begin{pmatrix} - (\tilde{f}_\kk^x - i \tilde{f}_\kk^y)^2& \cos\frac{\vartheta_\kk}2(\tilde{f}_\kk^x - i \tilde{f}_\kk^y) \\ -\cos\frac{\vartheta_\kk}2(\tilde{f}_\kk^x - i \tilde{f}_\kk^y) &  \cos^2\frac{\vartheta_\kk}2\end{pmatrix}\\
    &=&  \Delta_2 \begin{pmatrix} - \sin^2\frac{\vartheta_\kk}2(\hat{f}_\kk^x - i \hat{f}_\kk^y) & \frac12 \sin\vartheta_\kk \\ -\frac12 \sin\vartheta_\kk & \cos^2\frac{\vartheta_\kk}2(\hat{f}_\kk^x + i \hat{f}_\kk^y) \end{pmatrix}\\
&=&  \frac{\Delta_1}2 [\sin\vartheta_\kk \hat{\sigma}^0 + (\hat{f}_\kk^x \hat{\sigma}^x + \hat{f}_\kk^y\hat{\sigma}^y) - i \cos\vartheta_\kk(\hat{f}_\kk^x \hat{\sigma}^y - \hat{f}_\kk^y\hat{\sigma}^x)]i\hat{\sigma}^y.
    \label{eq:c2z_delta2_ss}
\end{eqnarray}
Finally, we find for the full gap function with time-reversal-symmetry breaking
\begin{equation}
    \hat{\Delta}(\kk) = \frac12[\sin\vartheta_\kk(\Delta_1 + \Delta_2)\hat{\sigma}_0 + (\Delta_2 - \Delta_1)(\hat{f}_\kk^x \hat{\sigma}^x + \hat{f}_\kk^y \hat{\sigma}^y) - i \cos\vartheta_\kk(\Delta_1 + \Delta_2)(\hat{f}_\kk^x \hat{\sigma}^y - \hat{f}_\kk^y \hat{\sigma}^x)](i\hat{\sigma}^y).
    \label{eq:c2v_full_delta}
\end{equation}
The last term in the above equation transforms as $A_{1u}$ under crystal symmetries and is odd under time reversal. This term is thus allowed for a system with $C_{4v}$ symmetry and time-reversal symmetry broken in $z$ direction (corresponding to $A^-_{2g}$). It is this term that breaks the balance of counter propagating currents for up and down spin and thus results in a net charge current at boundaries.

\subsection{$M_z \mathcal{T}$: TMD with an in-plane field}
We can use a similar single-band Hamiltonian as above as a toy model for a single layer of a transition-metal dichalcogenide with an in-plane magnetization, which can provide an example of a system lacking both inversion and time-reversal symmetry, but having $M_z \mathcal{T}$.
In particular, we can write
\begin{equation}
    \hat{\HH} = \epsilon_{\kk} + m_x \hat{\sigma}^x + m_y \hat{\sigma}^y +  f_{\kk}^z \hat{\sigma}^z
    \label{eq:tmd0}
\end{equation}
with $\epsilon_{\kk}$ now a lattice version of the diagonal energy and $f_{\kk}^{z} = - f_{-\kk}^z$. We find again similar transformation matrices
\begin{equation}
   \hat{U}_{\kk} = [\cos\frac{\vartheta_\kk}2\hat{\sigma}^0 - i \sin\frac{\vartheta_\kk}2(\hat{m}_y \hat{\sigma}^x -  \hat{m}_x \hat{\sigma}^y)] = \begin{pmatrix} \cos\frac{\vartheta_\kk}2 &  \sin\frac{\vartheta_{\kk}}2 e^{-i\phi}\\ - \sin\frac{\vartheta_\kk}2 e^{i\phi}   & \cos\frac{\vartheta_\kk}2\end{pmatrix},
    \label{eq:tmz_transformation}
\end{equation}
with $e^{i\phi} = \hat{m}_x + i \hat{m}_y$ and  $\hat{m}_i = m_i/\sqrt{m_x^2 + m_y^2}$. Note that now the angles are defined through 
\begin{equation}
    \sin\vartheta_\kk = \frac{\sqrt{m_x^2 + m_y^2}}{\sqrt{m_x^2 + m_y^2 + (f^z_{\kk})^2}} \qquad {\rm and} \qquad \cos\vartheta_\kk = \frac{f^z_{\kk}}{\sqrt{m_x^2 + m_y^2 + (f^z_{\kk})^2}}.
    \label{eq:tmz_angles}
\end{equation}
This leads to the following gap structure when transforming back:
\begin{eqnarray}
    \hat{\Delta}_1(\kk) &=&  U_{-\kk}^{\dag} \begin{pmatrix} \psi_1(\kk) & 0 \\ 0 & 0\end{pmatrix} U_{\kk}^*\\
    &=&  \psi_1(\kk)\begin{pmatrix} \sin\frac{\vartheta_\kk}2 &  -\cos\frac{\vartheta_{\kk}}2 e^{-i\phi} \\  \cos\frac{\vartheta_\kk}2e^{i\phi}  & \sin\frac{\vartheta_\kk}2\end{pmatrix}\begin{pmatrix} 1 & 0 \\ 0 & 0\end{pmatrix} \begin{pmatrix} \cos\frac{\vartheta_\kk}2 &  \sin\frac{\vartheta_{\kk}}2 e^{i\phi} \\ - \sin\frac{\vartheta_\kk}2 e^{-i\phi}  & \cos\frac{\vartheta_\kk}2\end{pmatrix}\\
    &=& \psi_1(\kk)\begin{pmatrix} \sin\frac{\vartheta_\kk}2 \cos\frac{\vartheta_\kk}2& \sin^2\frac{\vartheta_\kk}2 e^{i\phi} \\ \cos^2\frac{\vartheta_\kk}2 e^{i\phi} & \sin\frac{\vartheta_{\kk}}2\cos\frac{\vartheta_\kk}2 e^{2i\phi} \end{pmatrix}\\
    &=&  \frac{\tilde{\psi}(\kk)}2 \begin{pmatrix} \sin\vartheta_\kk e^{-i\phi} & 1 - \cos\vartheta_\kk \\ 1 + \cos\vartheta_\kk & \sin\vartheta_\kk e^{i\phi} \end{pmatrix}\\
    &=&  \frac{\tilde{\psi}_1(\kk)}2 [- \cos\vartheta_\kk \hat{\sigma}^0 + \hat{\sigma}^z + i \sin\vartheta_\kk (\hat{m}_y \hat{\sigma}^x - \hat{m}_x\hat{\sigma}^y)]i\hat{\sigma}^y,
    \label{eq:tmz_delta1_ss}
\end{eqnarray}
where we have absorbed the phase into $\tilde{\psi}_\kk$.
Using the simplest trivial function for $\tilde{\psi}_1(\kk) = \Delta_1 f_{\kk}^z$ and adding the same on the other band, we thus find for the case of $M_z\mathcal{T}$ conserved
\begin{equation}
    \hat{\Delta}(\kk) = \frac{f_{\kk}^z}2\Big[(\Delta_1 + \Delta_2)\cos\vartheta_\kk \hat{\sigma}^0 + (\Delta_1 - \Delta_2)\hat{\sigma}^z + i (\Delta_1 + \Delta_2)\sin\vartheta_\kk (\hat{m}_y \hat{\sigma}^x - \hat{m}_x\hat{\sigma}^y)\Big]i\hat{\sigma}^y.
    \label{eq:tmz_delta}
\end{equation}
In the simplest case, this describes an ($s$+$f$)-wave state that, due to the additional time-reversal-symmetry breaking, has an additional spin-triplet component. Note that the strength of the magnetic field can change the balance of the individual components and thus, the character of the order parameter potentially from dominant $s$-wave to dominant (nodal) $f$-wave.

\section{single-orbital model for FeSe}
In the following, we roughly follow the notation and discussion of \textcite{fischer:2011b}.
\subsection{Single-particle Hamiltonian}
\label{sec:singleparticle}
The staggered structure of FeAs can be incorporated in a single-orbital tight-binding description by defining the operatores
\begin{equation}
    c_{\alpha \kk s} = \left\{\begin{array}{ll} c_{\kk s} & \alpha = 1\\ c_{\kk + \Q s} & \alpha = 2\end{array}\right. ,
    \label{eq:twoband}
\end{equation}
where $\Q = (\pi, \pi)$ and we use $ \alpha = 1,2 $ as band indices. 
The Hamiltonian without the AFM order can then be written using Pauli matrices $\tau^i$ as
\begin{equation}
    \HH^0 = \sum_{\kk}\sum_{ss'}\sum_{\alpha\alpha'}(\epsilon_{\kk}^{0} \sigma^0_{ss'}\otimes\tau^0_{\alpha\alpha'} + \epsilon_{\kk}^{3}\sigma^0_{ss'}\otimes\tau^3_{\alpha\alpha'} + \vec{f}_{\kk}\cdot \vec{\sigma}_{ss'}\otimes\tau^1_{\alpha\alpha'})c_{\alpha \kk s}^{\dag}c_{\alpha'\kk s'}^{\phantom{\dag}},
    \label{eq:H0}
\end{equation}
with 
\begin{eqnarray}
    \epsilon_{\kk}^{0} &=& -4t'\cos k_x \cos k_y - \mu,\\
    \epsilon_{\kk}^{3} &=& -2t(\cos k_x + \cos k_y),\\
    f_{\kk}^{x} &=& \alpha \sin k_x \cos k_y,\label{eq:fx}\\
    f_{\kk}^{y} &=& - \alpha \cos k_x \sin k_y,\label{eq:fy}\\
    f_{\kk}^{z} &=&  0.
    \label{eq:mom0}
\end{eqnarray}
In the following, the sum over repeated indices is implied and when possible, the indices of the Pauli matrices are omitted. 
Note that we can see that this Hamiltonian has indeed the correct symmetry -- it belongs to the irreducible representation (IR) $A_{1g}$ -- from the transformation properties of the $\tau$ matrices (Table~\ref{tab:taus}) and the momentum functions (Table~\ref{tab:gapfunctions}).

Next, we include terms emerging when looking at an AFM order on the Fe sites with spins oriented in the $z$-direction. First, there is a Zeeman-like term reading
\begin{equation}
    \HH^{\rm AFM} = \sum_{\kk}m_z \sigma^z \otimes \tau^1c_{\alpha \kk s}^{\dag}c_{\alpha'\kk s'}^{\phantom{\dag}}.
    \label{eq:afm}
\end{equation}
Note that this term is odd under time-reversal symmetry and transforms as $B^-_{2u} = A_{2g}^{-} \otimes B_{1u}$ with $\pm$ referring to the behavior under TRS. There is then an additional term allowed,
\begin{equation}
    \HH' = \sum_{\kk} g_{\kk}\sigma^0 \otimes \tau^2c_{\alpha \kk s}^{\dag}c_{\alpha'\kk s'}^{\phantom{\dag}},
    \label{eq:AFMSOC}
\end{equation}
with a momentum dependence
\begin{equation}
    g_{\kk} = \sin k_x \sin k_y (\cos k_x - \cos k_y),
    \label{eq:gk}
\end{equation}
which corresponds to a 4th-nearest-neighbor hopping. This term can thus be expected to be small.

\begin{table}[bt]
    \centering
\begin{tabular}{c|cc|c}
     & intra-sublattice & inter-sublattice & IR\\
	 \hline
	 intra-band& $\tau^0$ & $\tau^3$ & $A_{1g}$\\
	 inter-band & $\tau^1$ & $\tau^{2}$ &$B_{1u}$
\end{tabular}
\caption{The different band dependencies possible for terms in the Hamiltonian of the systems under investigation here with the Pauli matrices $\tau$ acting in the space $\{\kk, \kk+\Q\}$.}
    \label{tab:taus}
\end{table}

It is convenient for the following to use the formulation by means of Green's functions, which for the non-interacting case can straightforwardly be calculated 
by inverting the $(4\times4)$ matrix $(i\omega_n \sigma^0\otimes\tau^0 -\HH_{\kk})$, with $\HH_{\kk} = \HH_{\kk}^0 + \HH_{\kk}^{\rm AFM} + \HH_{\kk}'$,
\begin{equation}
    \hat{G}_0(\kk, \omega_n) = G_{0+}(\kk, \omega_n)\sigma^0\otimes\tau^0 + G_{0-}(\kk, \omega_n)(\hat{f}_\kk\cdot\vec{\sigma}\otimes\tau^1 + \hat{m}_z \sigma^z\otimes\tau^1 + \hat{g}_{\kk}\sigma^0\otimes\tau^2+ \hat{\epsilon}_{\kk}\sigma^0\otimes\tau^3),
    \label{eq:greens0}
\end{equation}
where
\begin{equation}
    G_{0\pm}(\kk, \omega_n) = \frac{1}{2}\Big(\frac{1}{i\omega_n - \xi_{+, \kk}}\pm \frac{1}{i\omega_n - \xi_{-, \kk}}\Big)
    \label{eq:c1-greens0pm},
\end{equation}
\begin{eqnarray}
  \hat{f}^x_{\kk} &=& f_{\kk}^{x}/\sqrt{(f_\kk^{x})^2 + (f_\kk^{y})^2 +(m_{z})^2 +(g_\kk)^2 +(\epsilon_{\kk}^{3})^2},\\
  \hat{f}^y_{\kk} &=& f_{\kk}^{y}/\sqrt{(f_\kk^{x})^2 + (f_\kk^{y})^2 +(m_{z})^2 +(g_\kk)^2 +(\epsilon_{\kk}^{3})^2},\\
  \hat{f}^z_{\kk} &=& 0,\\
  \hat{m}_z &=&  m_z/\sqrt{(f_\kk^{x})^2 + (f_\kk^{y})^2 +(m_{z})^2 +(g_\kk)^2 +(\epsilon_{\kk}^{3})^2},\\
  \hat{g}_{\kk} &=&  g_{\kk}/\sqrt{(f_\kk^{x})^2 + (f_\kk^{y})^2 +(m_{z})^2 +(g_\kk)^2 +(\epsilon_{\kk}^{3})^2},
\end{eqnarray}
and 
\begin{equation}
  \hat{\epsilon}_{\kk} = \epsilon_{\kk}^{3}/\sqrt{(f_\kk^{x})^2 + (f_\kk^{y})^2 +(m_{z})^2 +(g_\kk)^2 +(\epsilon_{\kk}^{3})^2}.
\end{equation}
In Eq.~\eqref{eq:c1-greens0pm}, the two (doubly-degenarate) band energies are given by
\begin{equation}
  \xi_{\pm, \kk s} = \xi_{\pm, \kk} = \epsilon_{\kk}^{0} - \mu \pm\sqrt{(f_\kk^{x})^2 + (f_\kk^{y})^2 +(m_{z})^2 +(g_\kk)^2 +(\epsilon_{\kk}^{3})^2}. 
    \label{eq:sym-c1-energies0}
\end{equation}

\begin{table}[bt]
    \centering
\begin{tabular}{c|c|c}
  & \;\;\; intra-sublattice \;\;\;& \;\;\;inter-sublattice\;\;\;\\
	 \hline
	 \;\;$A_{1g}$ \;\;& $1$, $\cos k_x\cos k_y$ & $\cos k_x + \cos k_y$\\
	 $B_{1g}$ & - & $\cos k_x - \cos k_y$\\
	  $B_{2g}$ & $ \sin k_x \sin k_y $ & -\\ [0.5ex]
	 \hline
	 $A_{1u}$ & $\hat{x}\sin k_x \cos k_y + \hat{y}\sin k_y \cos k_x $ & $\hat{x}\sin k_x + \hat{y}\sin k_y$ \\
	 $A_{2u}$ & $\hat{y}\sin k_x \cos k_y - \hat{x}\sin k_y \cos k_x $ & $\hat{x}\sin k_y - \hat{y}\sin k_x$ \\
	 $B_{1u}$ & $\hat{x}\sin k_x \cos k_y - \hat{y}\sin k_y \cos k_x $ & $\hat{x}\sin k_x - \hat{y}\sin k_y$ \\
	 $B_{2u}$ & $\hat{y}\sin k_x \cos k_y + \hat{x}\sin k_y \cos k_x $ & $\hat{x}\sin k_y + \hat{y}\sin k_x$ \\
	  $ E_u $ & $ \{\hat{z} \sin k_x \cos k_y , \hat{z} \sin k_y \cos k_x \} $ & $ \{ \hat{z} \sin k_x , \hat{z} \sin k_y \} $
 \end{tabular}
\caption{Lowest order basis functions supported by intra- and inter-sublattice interactions on the lattice considered.}
    \label{tab:gapfunctions}
\end{table}

\subsection{Superconductivity}
We now turn to the problem of superconductivity by introducing a pairing interaction of the general form, 
\begin{equation}
    \HH'=\frac{1}{N}\sum_{\kk, \kk'}V_{\alpha\beta, \mu\nu}^{ss', s_3s_4}(\kk, \kk')c_{\alpha \kk s}^{\dag}c_{\beta-\kk s'}^{\dag}c_{\mu-\kk's_3}^{\phantom{\dag}}c_{\nu \kk's_4}^{\phantom{\dag}} .
    \label{eq:int}
\end{equation}
We parametrize the matrix element in the notation used for the single-particle terms,
\begin{equation}
  V_{\alpha\beta, \mu\nu}^{ss', s_3s_4}(\kk, \kk')=\sum_{m,n}\sum_{a}v_{mn}^{(a)}[\psi_{mn}^{(a)}(\kk)(\sigma^{m}i\sigma^y)_{ss'}\tau^{n}_{\alpha\beta}][\psi_{mn}^{(a)}(\kk')(\sigma^{m}i\sigma^y)_{s_{3}s_{4}}\tau^{n}_{\mu\nu}]^{\dag},
	\label{eq:c1-pairing}
\end{equation}
where $\psi_{mn}^{(a)}(\kk)$ have the symmetry of the gap functions tabulated in Tab.\ref{tab:gapfunctions}. 
The interaction matrix element $[V(\kk, \kk')]_{\alpha\beta\gamma\delta}^{s_1s_2s_3s_4}$ has an odd and an even part in $\kk$ which depends on the resulting sign of an interchange of the two first index pairs, $(\alpha\beta, s_1s_2) \leftrightarrow(\beta\alpha, s_2s_1)$,
\begin{equation}
  [V(\kk, \kk')]_{\alpha\beta\gamma\delta}^{s_1s_2s_3s_4} =v^{+}_{\kk \kk'}\Lambda_{+,\alpha\beta\gamma\delta}^{s_1s_2s_3s_4} + v^{-}_{\kk \kk'}\Lambda_{-,\alpha\beta\gamma\delta}^{s_1s_2s_3s_4},
    \label{eq:int-split}
\end{equation}
where 
\begin{equation}
  v^{\pm}_{\kk,\kk'} = \frac{1}{2}(v(\kk - \kk') \pm v(\kk + \kk')).\nonumber\\
    \label{eq:interplus}
\end{equation}
The corresponding matrix elements $\Lambda_{\pm, \alpha\beta\gamma\delta}^{s_1s_2s_3s_4}$ can be found in Ref.~\onlinecite{fischer:2011b}.

With the Hamiltonian and the non-interacting Green's function introduced above it is possible to analyze the superconducting instabilities in detail by resorting to the standard framework of the Gor'kov equations. The linearized gap equation reads
\begin{equation}
	\Delta_{\alpha\beta}^{ss'}(\kk)= -T  \sum_{\mu,\nu}\sum_{\omega_n}\sum_{\kk'} \sum_{s_3,s_4}  V_{\alpha\beta,\mu\nu}^{ss's_3s_4}(\kk, \kk')[\hat{G}_0(\kk', \omega_n) \hat{\Delta}(\kk') \hat{G}_0^T(-\kk', -\omega_n)]_{\nu\mu}^{s_4s_3}, 
    \label{eq:lingap}
\end{equation}
where all the Green's functions as well as the order parameter are $4\times4$ matrices. This gap equation is analyzed in the following for the two cases of a leading instability in the intra-sublattice and the inter-sublattice pairing channel, respectively.
For simplicity, we will start with a system without the AFM order, and then analyze how the stable solutions are suppressed, once AFM order is `turned on'.

\subsubsection{No magnetic order}
We look first at intra-sublattice pairing such that we have gap functions of the form
\begin{equation}
    \hat{\Delta}(\kk) = \left\{ \begin{array}{l} \psi_{0}(\kk)(i\sigma^y)\otimes \tau^{0} +  (\vec{d}_{1}(\kk)\cdot \vec{\sigma})(i\sigma^y) \otimes \tau^{1} \\
    \psi_{1}(\kk) (i\sigma^y) \otimes \tau^{1} + (\vec{d}_{0}(\kk)\cdot \vec{\sigma})(i\sigma^y) \otimes \tau^{0}
    \end{array} \right..
    \label{eq:intragap}
\end{equation}
The gap functions \eqref{eq:intragap} couple within the linearized gap equation (\ref{eq:lingap}),
\begin{equation}
  \psi_{0}(\kk) = -T\sum_{n,\kk'}v^{+}_{\kk \kk'}\Big\{[G_{0+}\tilde{G}_{0+} + G_{0-}\tilde{G}_{0-}]\psi_{0}(\kk') + [G_{0+}\tilde{G}_{0-}+G_{0-}\tilde{G}_{0+}]\hat{f}_{\kk'}\cdot\vec{d}_{1}(\kk')\Big\},\label{eq:c1-3d-singlet1}
\end{equation}
\begin{multline}
  \vec{d}_{1}(\kk) = -T\sum_{n,\kk'}v^{-}_{\kk \kk'}\Big\{[G_{0+}\tilde{G}_{0+} + G_{0-}\tilde{G}_{0-} ]\vec{d}_{1}(\kk')+2G_{0-}\tilde{G}_{0-} \{ \hat{f}_{\kk'} [\hat{f}_{\kk'}\cdot\vec{d}_{1}(\kk')] -\vec{d}_{1}(\kk')\} \\
     + [G_{0+}\tilde{G}_{0-}+G_{0-}\tilde{G}_{0+}]\hat{f}_{\kk'}\psi_{0}(\kk')\Big\}
    \label{eq:c1-3d-singlet}
\end{multline}
and, analogously,
\begin{multline}
    \vec{d}_{0}(\kk) = -T\sum_{n,\kk'}v^{-}_{\kk \kk'}\Big\{[G_{0+}\tilde{G}_{0+} + G_{0-}\tilde{G}_{0-}]\vec{d}_{0}(\kk') + 2G_{0-}\tilde{G}_{0-}\{\hat{f}_{\kk'} [\hat{f}_{\kk'}\cdot \vec{d}_{0}(\kk')] - |\hat{f}_{\kk'}|^2\vec{d}_{0}(\kk')\}\\
    +[G_{0+}\tilde{G}_{0-}+G_{0-}\tilde{G}_{0+}]\hat{f}_{\kk'}\psi_{1}(\kk')\Big\},
\end{multline}
\begin{multline}
  \psi_{1}(\kk) = -T\sum_{n,\kk'}v^{+}_{\kk \kk'}\Big\{[G_{0+}\tilde{G}_{0+} +G_{0-}\tilde{G}_{0-}]\psi_{1}(\kk') -2 (\hat{\epsilon}_{\kk'})^{2}G_{0-}\tilde{G}_{0-}\psi_{1}(\kk')\\
     + [G_{0+}\tilde{G}_{0-}+G_{0-}\tilde{G}_{0+}]\hat{f}_{\kk'}\cdot\vec{d}_{0}(\kk')\Big\}.
    \label{eq:c1-3d-triplet}
\end{multline}
Here, we have introduced the short notation $G_{0\pm} = G_{0\pm}(\kk, \omega_{n})$ and $\tilde{G}_{0\pm} = G_{0\pm}(-\kk, -\omega_n)$. 

As we are not interested here in the gap mixing, we only look in the following on the diagonal part and the resulting linearlized gap equation.
For the triplet gap functions, it is obvious that there is a main difference between $d$ vectors parallel or perpendicular to $\hat{f}_{\kk}$. For the case of a parallel $d$ vector, which is the most stable, we find the same linearized gap equations for the singlet and triplet cases that come with $\tau^0$, namely
\begin{equation}
    1 = -V\sum_{\kk'}\sum_{a=\pm}\frac{\lambda^2_0(\kk')}{2 \xi_{a,\kk'}}\tanh\left( \frac{\xi_{a,\kk'}}{2T} \right),
    \label{eq:fit0}
\end{equation}
where $\lambda_0(\kk')$ is a place holder for the momentum structure of the respective gap and interaction, e.g. $\lambda_0(\kk) = \psi_0(\kk)$.
For the gap functions combined with $\tau^1$, we find 
\begin{equation}
    1 = -V\sum_{\kk'}\sum_{a=\pm}\Big[\frac{1-\hat{\epsilon}^2_{\kk}}{2 \xi_{a,\kk'}} + \frac{\hat{\epsilon}^2_{\kk}}{2\epsilon_{\kk}^{0}}\Big]\lambda^2_1(\kk')\tanh\left( \frac{\xi_{a,\kk'}}{2T} \right).
    \label{eq:unfit1}
\end{equation}
As only the first term in the above sum has the divergence needed for the Cooper instability, we see that this term is suppressed. Finally, we can also look at the case of a $d$ vector perpendicular to $\hat{f}_{\kk}$ combined with a $\tau^{0}$. Here, we find the linearized gap equation
\begin{equation}
    1 = -V\sum_{\kk'}\sum_{a=\pm}\Big[\frac{1-\hat{f}^2_{\kk}}{2 \xi_{a,\kk'}} + \frac{\hat{f}^2_{\kk}}{2\epsilon_{\kk}^{0}}\Big]\lambda_{0,\perp}^2(\kk')\tanh\left( \frac{\xi_{a,\kk'}}{2T} \right).
    \label{eq:unfit0}
\end{equation}
Obviously, in this case, the suppression stems from the spin-orbit coupling parametrized by $\hat{f}_{\kk}$.


Turning to the inter-sublattice pairing, we only look at the most stable states to start with, i.e.,
\begin{equation}
 \hat{\Delta}(\kk) = \left\{ \begin{array}{l} (\vec{d}_{3}(\kk)\cdot \vec{\sigma})(i\sigma^y) \otimes \tau^{3} \\
    \psi_{3}(\kk) (i\sigma^{y}) \otimes \tau^{3}
    \end{array} \right..
    \label{eq:c1-intergap}
\end{equation}
To write the linearized gap equation we use the inter-sublattice pairing interactions to find 
\begin{equation}
  \psi_{3}(\kk) = -T\sum_{n,\kk'}v^{+}_{\kk \kk'} \left\{  [ G_{0+}\tilde{G}_{0+} +G_{0-}\tilde{G}_{0-}] \psi_{3}(\kk') - 2  G_{0-}\tilde{G}_{0-} \hat{f}^{2}_{\kk'} \psi_{3}(\kk') \right\}
    \label{eq:even-singlet}
\end{equation}
and, in the same way,
\begin{equation}
  \vec{d}_{3}(\kk) = -T\sum_{n,\kk'}v^{-}_{\kk \kk'}\Big\{[G_{0+}\tilde{G}_{0+} + G_{0-}\tilde{G}_{0-}]\vec{d}_{3}(\kk') - 2G_{0-}\tilde{G}_{0-}[\hat{f}_{\kk'}\cdot\vec{d}_{3}(\kk')]\hat{f}_{\kk'}\Big\}.
    \label{eq:odd-triplet}
\end{equation}
Interestingly, in this case only the spin-triplet gap function with a $d$ vector perpendicular to the $\hat{f}$ vector is most stable, again with a linearized self-consistency equation
\begin{equation}
    1 = -V\sum_{\kk'}\sum_{a=\pm}\frac{\lambda^2_{3,t}(\kk')}{2 \xi_{a,\kk'}}\tanh\left( \frac{\xi_{a,\kk'}}{2T} \right).
    \label{eq:fit3}
\end{equation}
Note that there are in principle two possibilities for a $d$ vector perpendicular to $\hat{f}$, namely one in plane and the other out of plane. These correspond to the gaps introduced in the main text in Eqs.~$(10)$ and $(11)$. However, the in-plane $d$ vector has a non-trivial momentum structure: in order for a vector to be perpendicular to $\hat{f}$, i.e., Eqs~\eqref{eq:fx} and \eqref{eq:fy}, it must have a structure that is of intra-sublattice form. The inter-sublattice gap with in-plane $d$ vector can thus only be approximately perpendicular to $\hat{f}$.

Finally, the almost stable spin-singlet gap now leads to 
\begin{equation}
    1 = -V\sum_{\kk'}\sum_{a=\pm}\Big[\frac{1-\hat{f}^2_{\kk}}{2 \xi_{a,\kk'}} + \frac{\hat{f}^2_{\kk}}{2\epsilon_{\kk}^{0}}\Big]\lambda^2_{3,s}(\kk')\tanh\left( \frac{\xi_{a,\kk'}}{2T} \right),
    \label{eq:unfit3}
\end{equation}
which is now suppressed by the spin-orbit coupling.

\subsubsection{AFM order}
For the case of AFM ordered FeSe, we only focus on the (almost) stable states found above and look at their `diagonal' selfconsistency equation, i.e.,
\begin{equation}
    \hat{\Delta}(\kk) = \left\{ \begin{array}{l} \psi_{0}(\kk) (i\sigma^y)\otimes \tau^{0}\\
    (\vec{d}_{0}(\kk)\cdot \vec{\sigma})(i\sigma^y) \otimes \tau^{0}
    \end{array} \right.,
    \label{eq:intragap_AFM}
\end{equation}
with $d_0(\kk)||\hat{f}_{\kk}$ for the intra-sublattice pairing states and
\begin{equation}
 \hat{\Delta}(\kk)= \left\{ \begin{array}{l} (\vec{d}_{3}(\kk)\cdot \vec{\sigma})(i\sigma^y) \otimes \tau^{3} \\
    \psi_{3}(\kk) (i\sigma^y) \otimes \tau^{3}
    \end{array} \right.,
    \label{eq:intergap_AFM}
\end{equation}
with $d_{3}(\kk)\perp \hat{f}_{\kk}$ for the inter-sublattice states. Again starting with the intra-sublattice orders, we find
\begin{equation}
    \psi_{0}(\kk) = -T\sum_{n,\kk'}v^{+}_{\kk \kk'}\Big\{[G_{0+}\tilde{G}_{0+} + G_{0-}\tilde{G}_{0-}]\psi_{0}(\kk') - 2 G_{0-}\tilde{G}_{0-}[\hat{g}^{2}_{\kk} + \hat{m}^2_z] \psi_{0}(\kk')\Big\},
  \label{eq:afm-singlet0}
\end{equation}
for the singlet, and
\begin{eqnarray}
    \vec{d}_{0}(\kk) &=& -T\sum_{n,\kk'}v^{-}_{\kk \kk'}\Big\{[G_{0+}\tilde{G}_{0+} + G_{0-}\tilde{G}_{0-}]\vec{d}_{0}(\kk') + 2G_{0-}\tilde{G}_{0-}\{\hat{f}_{\kk'} [\hat{f}_{\kk'}\cdot \vec{d}_{0}(\kk')] - (|\hat{f}_{\kk'}|^2 + \hat{g}_{\kk}^2)\vec{d}_{0}(\kk')\}\\
    &=&  -T\sum_{n,\kk'}v^{-}_{\kk \kk'}\Big\{[G_{0+}\tilde{G}_{0+} + G_{0-}\tilde{G}_{0-}]\vec{d}_{0}(\kk') - 2G_{0-}\tilde{G}_{0-}\hat{g}_{\kk'}^2\vec{d}_{0}(\kk')\},
\end{eqnarray}
where for the second line we have used $\vec{d}_{0}(\kk) \parallel \hat{f}_{\kk}$.

For the case of inter-sublattice pairing, we find for the spin-singlet gap
\begin{equation}
  \psi_{3}(\kk) = -T\sum_{n,\kk'}v^{+}_{\kk \kk'} \left\{  [ G_{0+}\tilde{G}_{0+} +G_{0-}\tilde{G}_{0-}] \psi_{3}(\kk') - 2  G_{0-}\tilde{G}_{0-} \hat{f}^{2}_{\kk'} \psi_{3}(\kk') \right\}
    \label{eq:afm-singlet3}
\end{equation}
which again is suppressed by the spin-orbit coupling. Finally, for the spin-triplet gap, we find
\begin{equation}
    \vec{d}^{\,\perp}_{3}(\kk) = -T\sum_{n,\kk'}v^{-}_{\kk \kk'}\Big\{[G_{0+}\tilde{G}_{0+} + G_{0-}\tilde{G}_{0-}]\vec{d}^{\,\perp}_{3}(\kk') -2G_{0-}\tilde{G}_{0-}\hat{m}_z^2\vec{d}^{\,\perp}_{3}(\kk')\Big\},
    \label{eq:afm-triplet-inplane}
\end{equation}
for an in-plane $d$ vector with $\vec{d}_{3}(\kk)\perp \hat{f}_{\kk}$ while 
\begin{equation}
  d^z_{3}(\kk) = -T\sum_{n,\kk'}v^{-}_{\kk \kk'}[G_{0+}\tilde{G}_{0+} + G_{0-}\tilde{G}_{0-}]d^z_{3}(\kk')
    \label{eq:afm-triplet-z}
\end{equation}
for the $d$ vector along the $z$ direction.

Finally, we can again look at the linearized self-consistency equations determining the critical temperature after summing over the Matsubara frequencies to find
\begin{equation}
    1 = -V\sum_{\kk'}\sum_{a=\pm}\Big[\frac{1-(\hat{g}^2_{\kk} + \hat{m}_{z}^2)}{2 \xi_{a,\kk'}} + \frac{\hat{g}^2_{\kk} + \hat{m}_{z}^2}{2\epsilon_{\kk}^{0}}\Big]\lambda_{0,s}^2(\kk')\tanh\left( \frac{\xi_{a,\kk'}}{2T} \right)
    \label{eq:afm_unfit_s0}
\end{equation}
for the intra-sublattice singlet,
\begin{equation}
    1 = -V\sum_{\kk'}\sum_{a=\pm}\Big[\frac{1-\hat{g}^2_{\kk}}{2 \xi_{a,\kk'}} + \frac{\hat{g}^2_{\kk}}{2\epsilon_{\kk}^{0}}\Big]\lambda_{0,t}^2(\kk')\tanh\left( \frac{\xi_{a,\kk'}}{2T} \right)
    \label{eq:afm_unfit_t0}
\end{equation}
for the intra-sublattice triplet,
\begin{equation}
    1 = -V\sum_{\kk'}\sum_{a=\pm}\Big[\frac{1-\hat{f}^2_{\kk}}{2 \xi_{a,\kk'}} + \frac{\hat{f}^2_{\kk}}{2\epsilon_{\kk}^{0}}\Big]\lambda^2_{3,s}(\kk')\tanh\left( \frac{\xi_{a,\kk'}}{2T} \right)
    \label{eq:afm_unfit_s3}
\end{equation}
for the inter-sublattice singlet, and finally
\begin{equation}
    1 = -V\sum_{\kk'}\sum_{a=\pm}\frac{\lambda^2_{3,t}(\kk')}{2 \xi_{a,\kk'}}\tanh\left( \frac{\xi_{a,\kk'}}{2T} \right)
    \label{eq:afm_fit_t3}
\end{equation}
for the inter-sublattice triplet with a $d$ vector parallel $m_z$.
The most stable gap is thus the inter-sublattice triplet state with a $d$ vector parallel to the magnetic order, i.e., in $z$ direction [Eq.~\eqref{eq:afm_fit_t3}]. Note, however, that the intra-sublattice triplet state with $d\parallel \vec{f}$ is almost as stable [Eq.~\eqref{eq:afm_unfit_t0}], and the inter-sublattice singlet state is also only suppressed by the spin-orbit coupling [Eq.~\eqref{eq:afm_unfit_s3}]. This might or might not be a strong term (what counts is the relative strength compared to the nearest-neighbor hopping).

\subsubsection{Order parameter mixing and charge currents}
Finally, we want to look at order parameter (or gap) mixing for the case of the more stable gap functions above. In general, a gap belonging to an IR $R_1$ can couple to a gap of IR $R_2$, if $R^{\pm}_2 \in R_1^{\mp} \otimes B^{-}_{2u}$, with $B^{-}_{2u} = A^{-}_{2g} \otimes B^{+}_{1u}$ the symmetry of the staggered magnetization and the superscript $\pm$ refering to the behavior under TRS.

The most stable order parameter $d_{3}^{z}(\kk)\sigma^z (i\sigma^y)\otimes \tau^3$ belongs to $E_u$ and we do not find any interesting coupling. However, for the almost stable states we find gap mixing similar to what we found for the Rashba case. First, for the spin-singlet case with overall $A_{1g}$ symmetry, there are two spin-triplet gap functions that can be mixed in, namely 
\begin{equation}
    \Delta_{A_{1g}}^{(1)}(\kk) = \Delta_1(\sin k_x \cos k_y \sigma^x - \sin k_y \cos k_x \sigma^y) (i\sigma^y) \otimes\tau^1
\end{equation}
of $B^+_{1u}\otimes B^+_{1u}$ symmetry, which is allowed to mix even in the absence of magnetic order [see Eq.~\eqref{eq:c1-3d-singlet1}], and 
\begin{equation}
    \Delta_{A_{1g}}^{(2)}(\kk) = i \Delta_2 (\sin k_y \cos k_x \sigma^x + \sin k_x \cos k_y \sigma^y) (i\sigma^y)\otimes\tau^0
    \label{eq:dp2}
\end{equation}
of $B^-_{2u} \times A^+_{1g}$ symmetry. Going to sublattice (instead of band) space, $\tau^1$ becomes $\tau^3$, such that on one sublattice
\begin{equation}
    \Delta^A_{A_{1g}}(\kk) = \begin{pmatrix}
	- (\Delta_1 + \Delta_2)(\sin k_x \cos k_y + i \sin k_y \cos k_x) & 0 \\
	0 & (\Delta_1 - \Delta_2)(\sin k_x \cos k_y - i \sin k_y \cos k_x) 
    \end{pmatrix}
    \label{eq:subA1}
\end{equation}
while on the other
\begin{equation}
    \Delta^B_{A_{1g}}(\kk) = \begin{pmatrix}
	(\Delta_1 - \Delta_2)(\sin k_x \cos k_y + i \sin k_y \cos k_x) & 0 \\
	0 & - (\Delta_1 + \Delta_2)(\sin k_x \cos k_y - i \sin k_y \cos k_x) 
    \end{pmatrix}.
    \label{eq:subB1}
\end{equation}
Similarly, for the case of a spin-triplet gap of the form
\begin{equation}
    \Delta_{B_{1u}}^{(1)}(\kk) = \Delta_1(\sin k_x \cos k_y \sigma^x - \sin k_y \cos k_x \sigma^y) (i\sigma^y) \otimes\tau^0
    \label{eq:b1u}
\end{equation}
the AFM order leads to a mixing with
\begin{equation}
    \Delta_{B_{1u}}^{(2)}(\kk) = i \Delta_2 (\sin k_y \cos k_x \sigma^x + \sin k_x \cos k_y \sigma^y) (i\sigma^y)\otimes\tau^1
    \label{eq:b1ut}
\end{equation}
which leads to
\begin{equation}
    \Delta^A_{B_{1u}}(\kk) = \begin{pmatrix}
	- (\Delta_1 + \Delta_2)(\sin k_x \cos k_y + i \sin k_y \cos k_x) & 0 \\
	0 & (\Delta_1 - \Delta_2)(\sin k_x \cos k_y - i \sin k_y \cos k_x) 
    \end{pmatrix}
    \label{eq:subA2}
\end{equation}
and
\begin{equation}
    \Delta^B_{B_{1u}}(\kk) = \begin{pmatrix}
	-(\Delta_1 - \Delta_2)(\sin k_x \cos k_y + i \sin k_y \cos k_x) & 0 \\
	0 & (\Delta_1 + \Delta_2)(\sin k_x \cos k_y - i \sin k_y \cos k_x) 
    \end{pmatrix}.
    \label{eq:subB2}
\end{equation}
In both cases, there is thus a $p+ip$ order parameter for spin up and a $p-ip$ for spin down with an imbalance that is opposite on the two sublattices, resulting in no net current at a random boundary. However, for the right termination, e.g., a perfectrly diagonal boundary, there could be a net current.

\bibliography{B_refs}